\documentclass[10pt,fleqn]{article}
\pdfoutput=1
\usepackage{amsmath}
\usepackage{amssymb}
\usepackage{array}
\usepackage{calc}
\usepackage{longtable}
\usepackage{multirow,booktabs}
\usepackage{relsize}
\usepackage{pstricks}
\usepackage{graphicx}
\usepackage{xspace}
\usepackage{siunitx}

\numberwithin{equation}{section}
\usepackage{mciteplus}
\usepackage{tikz-feynman}
\usepackage[a4paper,pdfborder={0 0 0}]{hyperref}
\usepackage[format=hang,labelfont=bf,hypcap=true]{caption}
\usepackage{subcaption}
\usepackage{sectsty}
\usepackage{transparent}
\usepackage{setspace}
\allsectionsfont{\normalfont\sffamily\bfseries}
\subsectionfont{\normalfont\sffamily\bfseries}
\subsubsectionfont{\normalfont\sffamily\bfseries}
\makeatletter
\g@addto@macro\bfseries{\boldmath}
\makeatother

\let\spreprint\empty
\newcommand{\preprint}[1]{\def\spreprint{\protect#1}}
\let\sinstitute\empty
\newcommand{\institute}[1]{\def\sinstitute{\protect#1}}
\makeatletter
\renewcommand{\maketitle}{\begingroup
  \null\thispagestyle{empty}%
    \ifx\spreprint\empty
      \vskip 5ex
    \else
      \flushright\large\spreprint\vskip 2ex
    \fi
    \vskip 5ex
    \flushleft
      {\sffamily\bfseries\huge\@title}\vskip 6ex
      \@author\vskip 2ex
      \ifx\sinstitute\empty
      \else
        {\small\sinstitute}
      \fi
    \vskip 5ex
  \endgroup
}
\makeatother
\renewenvironment{abstract}{\begin{center}
  {\large\sffamily\bfseries Abstract: }
  \begin{minipage}[t]{0.75\textwidth}
}{\end{minipage}\end{center}\vskip 10ex}

\numberwithin{equation}{section}

\newcommand{\SHERPA}{\texorpdfstring{S\protect\scalebox{0.8}{HERPA}}{SHERPA}\xspace}
\newcommand{\Sherpa}{\SHERPA}

\newcommand{\COMIX}{C\protect\scalebox{0.8}{OMIX}\xspace}
\newcommand{\Comix}{\COMIX}
\newcommand{\Recola}{R\protect\scalebox{0.8}{ECOLA}\xspace}

\newcommand{\POWHEG}{P\protect\scalebox{0.8}{OWHEG}-B\protect\scalebox{0.8}{OX}\xspace}
\newcommand{\mgfive}{M\protect\scalebox{0.8}{AD}G\protect\scalebox{0.8}{RAPH}5\_\protect\scalebox{0.8}{A}MC@NLO\xspace}
\newcommand{\LHC}{L\protect\scalebox{0.8}{HC}\xspace}
\newcommand{\Fastjet}{F\protect\scalebox{0.8}{AST}J\protect\scalebox{0.8}{ET}\xspace}
\newcommand{\HepMC}{H\protect\scalebox{0.8}{EP}MC\xspace}

\newcommand{\MEPSatNLO}{M\protect\scalebox{0.8}{EPS@}N\protect\scalebox{0.8}{LO}\xspace}
\newcommand{\Alpgen}{A\protect\scalebox{0.8}{LP}G\protect\scalebox{0.8}{EN}\xspace}
\newcommand{\EWvirt}{EW$_\text{virt}$\xspace}

\hypersetup{
  pdfauthor={}
  pdftitle={}
}
\preprint{MCNET-20-12}
\author{Enrico Bothmann$^1$, Davide Napoletano$^{2}$}
\title{Automated evaluation of electroweak Sudakov \\[0.4em] logarithms in \Sherpa}
\institute{
  $^1$ \textit{Institut f\"ur Theoretische Physik, Georg-August-Universit\"at
  G\"ottingen, D-37077 G\"ottingen, Germany}\\
  $^2$ \textit{Universit\`a degli Studi di Milano-Bicocca \& INFN, Piazza della
  Scienza 3, Milano 20126, Italy}}
\begin{document}
\maketitle
\begin{abstract}
  We present an automated implementation for the calculation of one-loop double
  and single Sudakov logarithms stemming from electroweak radiative corrections
  within the \Sherpa event generation framework, based on the derivation
  in~\cite{Denner:2000jv}.
  At high energies, these logarithms constitute the leading
  contributions to the full NLO electroweak corrections. As examples,
  we show applications for relevant processes at both the \LHC and
  future hadron colliders,
  namely on-shell W boson pair production, EW-induced dijet production
  and electron-positron production in association with four jets,
  providing the first estimate of EW corrections at this multiplicity.
\end{abstract}

\section{Introduction}

In perturbative electroweak (EW) theory, higher-order corrections
are given by emissions of virtual and real gauge bosons and are known
to have a large effect in the hard tail of observables at the \LHC and
future colliders~\cite{Mishra:2013una}.
In contrast to massless gauge theories, where real
emission terms are necessary to regulate infrared divergences, in EW
theory the weak gauge bosons are massive and therefore 
provide a natural lower scale cut-off, such that finite logarithms
appear instead. Moreover, since the emission of an additional weak
boson leads to an experimentally different signature with respect to
the Born process, virtual logarithms are of physical significance
without the inclusion of real radiation terms.

The structure of such logarithmic contributions, referred to as
Sudakov logarithms~\cite{Sudakov:1954sw}, and their
factorisation properties were derived in full 
generality by Denner and
Pozzorini~\cite{Denner:2000jv,Denner:2002gd,Denner:2001gw,Denner:2004iz}
at leading and next-to-leading logarithmic accuracy at both one- and
two-loop order in the EW coupling expansion.
In particular, they have shown that in the high-energy scattering
limit, at least at one-loop order, these logarithmic 
corrections can be factored as a sum over pairs of external legs in an otherwise
process-independent way, hence providing a straightforward algorithm for
computing them for any given process. The high-energy limit
requires all invariants formed by
pairs of particles to be large compared to the scale set by the weak
gauge boson masses. In this sense we expect effects due to electroweak
corrections in general, and to Sudakov logarithms in particular, to
give rise to large effects in the hard tail of observables
that have the dimension of energy,
such as for example the transverse momentum of
a final state particle.

In this work we present an implementation for
computing EW Sudakov corrections in the form they are presented
in~\cite{Denner:2000jv} within the \Sherpa event generator
framework~\cite{Bothmann:2019yzt}.
While Sudakov corrections have been computed for a variety of processes
(e.g.\ \cite{Kuhn:2004em,Kuhn:2005gv,Accomando:2006hq,Kuhn:2007cv,Chiesa:2013yma}),
a complete and public implementation is, to our knowledge,
missing.\footnote{An implementation is referred to be available
  in~\cite{Chiesa:2013yma}. However, contrary to the one presented
  here it has two limitations: it
  is only available for the processes hard-coded in \Alpgen, and it
  does not include corrections for longitudinal external gauge bosons.}
The one we present here is fully general and automated, it is only limited by
computing resources, and it will be made public in the upcoming major release
of \Sherpa.
It relies on the internal matrix element
generator \Comix~\cite{Gleisberg:2008fv} to compute all double and
single logarithmic contributions at one-loop order for any user-specified
Standard Model process. 
The correction is calculated for each event individually, and is therefore
fully differential in phase-space, such that the logarithmic corrections for
any observable are available from the generated event sample.
The event-wise correction factors are written out as additional
weights~\cite{Dobbs:2001ck} along with the main event weight, such that
complete freedom is left to the user on how to use and combine these Sudakov
correction weights.

As Sudakov EW logarithms are given by the high-energy limit of the full set of
virtual EW corrections, they can approximate these effects.
Indeed, while NLO EW corrections are now becoming a standard in all general purpose
Monte Carlo event generators (see~\cite{Schonherr:2017qcj} for \Sherpa,
\cite{Frederix:2018nkq} for \mgfive and process-specific implementations for
\POWHEG~\cite{Alioli:2010xd}, such as e.g.\
\cite{Jager:2011ms,Bernaciak:2012hj,Barze:2012tt}), the full set of NLO EW
corrections for complicated final states may not
be available yet or might not be numerically tractable due to the
calculational complexity for a given application.
For these reasons, an approximate scheme to deal with NLO EW virtual
corrections, dubbed \EWvirt, was devised in~\cite{Kallweit:2015dum}.
Within this approximation NLO EW real emission terms are neglected,
and one only includes virtual contributions with the minimal amount
of counter-terms to be rendered finite. This approach greatly
simplifies the integration of such corrections with more complex event
generation schemes, such as matching and merging with a QCD parton shower and
higher-order QCD matrix elements. Compared to the Sudakov EW
corrections used as an approximation of the virtual NLO EW, the
\EWvirt scheme differs due to the inclusion of exact
virtual corrections and integrated subtraction
terms. These differences are either subleading or
constant at high energy in the logarithmic counting,
but they can lead to numerically sizeable effects,
depending on the process, the observable and the applied cuts.

Alternatively, one can exploit the fact that EW Sudakov logarithms have been
shown to exponentiate up to next-to-leading logarithmic accuracy using IR
evolution
equation~\cite{Fadin:1999bq,Melles:2001ye,Melles:2001dh,Jantzen:2005xi} and
using Soft-Collinear Effective
Theory~\cite{Chiu:2007yn,Chiu:2007dg,Chiu:2008vv,Chiu:2009mg,Chiu:2009ft,Fuhrer:2010eu},
and exponentiate the event-wise correction factor 
to calculate the fully differential next-to-leading logarithmic resummation,
as explored e.g.\ in~\cite{Lindert:2017olm}.\footnote{The fact the
all-order resummation, at NLL accuracy, can be obtained by exponentiating the
NLO result was first derived by Yennie, Frautschi and
Suura (YFS)~\cite{Yennie:1961ad} in the case of a massless gauge
boson in abelian gauge theories.}
This can lead to particularly important effects e.g.\ at a
\SI{100}{\TeV} proton-proton collider~\cite{Manohar:2014vxa},
where (unresummed) next-to-leading logarithmic corrections can approach
O(1) at the high-energy tail due to the increased kinematic reach,
hence even requiring resummation to obtain a valid prediction.

The outline of this paper is as follows.
In Sec.~\ref{sec:imple}, we present how the algorithm is implemented in the
\Sherpa event generator, while in Sec.~\ref{sec:examples-testin} we
show a selection of applications of the method used to test our implementation,
namely on-shell W boson pair production, EW-induced dijet production,
and electron-positron production in association with four jets,
providing an estimate of EW corrections for the latter process
for the first time and at the same time
proving the viability of the method for high multiplicity final states.
Finally we report our conclusions~in Sec.~\ref{sec:conclusions}.

\section{\Sherpa implementation}
\label{sec:imple}

Electroweak Sudakov logarithms arise from the infrared region of the
integration
over the loop momentum of a virtual gauge
boson exchange, with the lower integration boundary set by the gauge boson mass.
If the boson mass is small compared to the process energy scales,
they exhibit the same structure of soft and
collinear divergences one encounters in massless gauge theories. As
such, the most dominant contributions are given by double logarithms
(DL)
which arise from the soft and collinear limit, and single logarithms 
(SL) coming instead from less singular regions,
as we describe in more detail in the rest of this section. Virtual
EW corrections also contain in general non-logarithmic terms (either
constant or power suppressed), which are however beyond the accuracy
of this work.
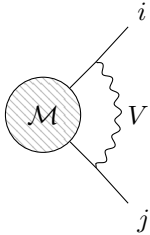
\begin{figure}[b]
\begin{center}
\begin{tikzpicture}
\begin{feynman}
  \vertex [large, pattern color=white!75!black, blob] (a) {$\mathcal M$};
  \vertex [node distance=1.0, above right=of a] (b);
  \vertex [node distance=1.0, below right=of a] (c);
  \vertex [node distance=0.6, above right=of b] (d) {$i$};
  \vertex [node distance=0.6, below right=of c] (e) {$j$};
	\diagram* {
    (a) -- (b),
    (a) -- (c),
    (b) -- (d),
    (c) -- (e),
    (b) -- [quarter left, photon, edge label=$V$] (c),
  };
\end{feynman}
\end{tikzpicture}
\caption{\label{fig:vloop}
Feynman diagram contributing to double-logarithmic Sudakov corrections.}
\end{center}
\end{figure}%
To give a more explicit example, let us consider the DL and SL contributions
coming from a diagram of the type shown in Fig.~\ref{fig:vloop}.
The former correspond to the case of the virtual gauge boson $V$
becoming soft and collinear to one of the external legs~$i$ or~$j$
of the matrix element $\mathcal M$, i.e.\ when
$|r_{ij}|\equiv|p_i+p_j|^2\gg m_W^2$.
Similarly, as $|p_i+p_j|^2\sim2\,p_i\cdot\,p_j$, this limit also encodes
the case of a collinear, non-soft (and vice-versa) gauge boson which
instead gives rise to SL contributions. In turn, these
must also be included, since they can numerically be of a
similar size as the DL one, see the discussion
and the detailed derivation
in~\cite{Denner:2000jv,Denner:2001gw}. The reason for this can also be
understood intuitively: while the diagram in Fig.~\ref{fig:vloop} is the only source of double
logarithms there are several contributions giving rise to single
logarithms, which we discuss later in this section. 
In all cases, following the conventions in~\cite{Denner:2000jv}, we write DL
and SL terms in the following way, respectively:
\begin{align}
  L\left(\left|r_{i j}\right|\right)
  \equiv\frac{\alpha}{4 \pi} \log ^{2} \frac{|r_{i j}|}{m_W^{2}}
  ,\quad
  l\left(r_{i j}\right)
  \equiv\frac{\alpha}{4 \pi} \log \frac{r_{i j}}{m_W^{2}}
  .
\end{align}
Note that the use of the $W$ mass as a reference scale is only a
convention and the same form is indeed also used for contributions coming
from $Z$ bosons and photons. Remainder logarithms containing ratios
of $m_W$ and $m_Z$ or the vanishing photon mass
are taken care of by additional logarithmic terms, as explained in
more details below.

After having evaluated all relevant diagrams, and using Ward identities,
one can show that the DL and SL contributions factorise off the Born matrix
element $\mathcal M$, and can be written in terms of a sum over pairs of
external legs only.
Hence, we define these corrections as a multiplicative factor~$K_{\text{NLL}}$
to the squared matrix element $|\mathcal M(\Phi)|^2$
at a given phase space point~$\Phi$:
\begin{equation}
  \label{eq:kfactor}
  K_{\text{NLL}}\left(\Phi\right)
  = 1 + \sum_c \frac{\sum_h 2\,{\rm Re}\left\{
      (\delta^c \mathcal M_h) \mathcal M_h^*
  \right\}}{\sum_h \left| \mathcal M_h \right|^2},
\end{equation}
where we have written $\mathcal M_h \equiv \mathcal M_h(\Phi)$
and the sums go over helicity configurations $h$
and the different types of logarithmic contributions $c$. Following
the original notation, these contributions are divided into Leading and
Subleading Soft and Collinear (LSC/SSC) logarithms, Collinear or Soft (C)
logarithms and those arising from Parameter Renormalisation (PR). More details
about their respective definitions are given below.

Since $\delta^c$ is in general a tensor in $\text{SU}(2)\times\text{U}(1)$, the
$\delta^c \mathcal M_h$ are usually not proportional to the original matrix
element $\mathcal M_h$.
The tensor structure come from the fact that in general the various
$\delta^c$ are proportional to EW vertices, which in turn means that a
single leg or pairs of legs can get replaced with particles of a
different kind, according to the relevant Feynman rules.
As in~\cite{Denner:2000jv}, we
denote the leg index $i$ as a shorthand for the external fields
$\phi_{i}$. Denoting with $\{i\}$ the set of all external fields,
we therefore have $\delta^c \mathcal M_h^{\{i\}} \propto \mathcal M_h^{\{i'\}}$.
In our implementation, the construction and evaluation of these
additional amplitudes is taken care of by an interface to the automated tree-level \Comix matrix element
generator~\cite{Gleisberg:2008fv}, which is available within the
\Sherpa framework.
Before evaluating such amplitudes, the energy is re-distributed among the
particles to put all (replaced) external legs on-shell again.
The required auxiliary processes are automatically set up during
the initialisation of the framework.
Since the construction of additional processes can be computationally non-trivial, we have
taken care of reducing their number by re-using processes that are shared between
different contributions $c$, and by using crossing relations when only the order
of external fields differs.

In our implementation we consistently neglect purely
electromagnetic logarithms which arise from the gap between zero (or
the fictitious photon mass) and $m_W$. In \Sherpa such
logarithms can be treated in either of two ways. First, one can
compute the purely electromagnetic NLO correction to the desired
process, consisting of both virtual and real photon emission, which
gives the desired logarithm at fixed order. Alternatively, one can resum
soft-photon emission to all orders, which in \Sherpa is achieved
through the formalism of Yennie, Frautschi and Suura
(YFS)~\cite{Yennie:1961ad}
or by a QED parton shower, as is discussed e.g.\ in~\cite{Kallweit:2017khh}.
In contrast, logarithms originating from the
running between the $W$ mass and the $Z$ mass are included, as we
explain in the next paragraphs, where we discuss the individual
contributions $\delta^c$.

\paragraph{Leading Soft-Collinear logarithms: LSC}
The leading corrections are given by the soft and collinear limit of a
virtual gauge boson and are proportional to $L(|r_{kl}|)$.
By writing
\begin{equation}\label{eq:DL}
  L\left(\left|r_{k l}\right|\right)=L\left(s\right)+2\,l(s) \log \frac{\left|r_{k l}\right|}{s}+L\left(\left|r_{k l}\right|, s\right),
\end{equation}
and neglecting the last term on the right-hand side,
which is of order $\mathcal{O}(1)$, one has split
the soft-collinear contribution into
a leading soft-collinear (LSC) and a angular-dependent subleading soft-collinear (SSC)
one, corresponding to the first and second term on the right-hand side, respectively.
The full LSC correction is then given by
\begin{align}\label{eq:LSC}
  \delta^{\mathrm{LSC}} \mathcal{M}^{i_{1} \ldots i_{n}}
  =\sum_{k=1}^{n} \delta_{i_{k}^{\prime} i_{k}}^{\mathrm{LSC}}(k)
  \mathcal{M}_{0}^{i_{1} \ldots i_{k}^{\prime} \ldots i_{n}}
,\end{align}
where the sum runs over the $n$ external legs, and the coefficient
matrix on the right-hand side is given by
\begin{align}
  \label{eq:LSC_coeff}
  \delta_{i_{k}^{\prime} i_{k}}^{\mathrm{LSC}}(k)
  =-\frac{1}{2}\left[C_{i_{k}^{\prime} i_{k}}^{\mathrm{ew}}(k)
    L(s)-2\left(I^{Z}(k)\right)_{i_{k}^{\prime} i_{k}}^{2} \log
    \frac{M_Z^{2}}{M_W^{2}} l(s)
  \right]
  = \delta_{i_{k}^{\prime} i_{k}}^{\overline{\mathrm{LSC}}}(k)
  + \delta_{i_{k}^{\prime} i_{k}}^Z(k).
\end{align}
$C^\text{ew}$ and $I^Z$ are the electroweak Casimir operator and
the $Z$ gauge coupling, respectively.
The second term $\delta^Z$ appears as an artefact of writing $L(s)$ and $l(s)$
in terms of the $W$ mass even for $Z$ boson loops, and hence the inclusion of
this term takes care of the gap between the $Z$ and the $W$ mass at
NLL accuracy.
We denote the remaining terms with the superscript ``$\overline{\text{LSC}}$''.
Note that this term is in general non-diagonal, since $C^\text{ew}$ mixes
photons and $Z$ bosons.
As is explicit in Eq.~\eqref{eq:kfactor}, coefficients need to be computed per
helicity configuration;
in the case of longitudinally polarised vector boson appearing as external
particles, they need to be replaced with the corresponding Goldstone bosons
using the Goldstone Boson Equivalence Theorem.
This is a consequence of the scaling of longitudinal polarisation vectors with
the mass of the particle, which prevents the direct application of the eikonal
approximation used to evaluate the high-energy limit, as is detailed
in~\cite{Denner:2000jv}.
To calculate such
matrix elements we have extended the default Standard Model
implementation in \Sherpa with a full Standard Model including
Goldstone bosons generated through the UFO
interface of \Sherpa~\cite{Degrande:2011ua,Hoche:2014kca}. We have tested this
implementation thoroughly against
\Recola~\cite{Uccirati:2014fda}.\footnote{Note that in particular this adds the
  possibility for \Sherpa users to compute Goldstone bosons contributions to
  any desired process, independently of whether this is done in the context
  of calculating EW Sudakov corrections.}

\paragraph{\bf Subleading Soft-Collinear logarithms: SSC}
The second term in Eq.~\ref{eq:DL} gives rise to the angular-dependent
subleading part of the corrections from soft-collinear gauge-boson loops.
It can be written as a sum over pairs of external particles:
\begin{align}\label{eq:deltaSSC}
  \delta^{\mathrm{SSC}} \mathcal{M}^{i_{1} \ldots i_{n}}
  =\sum_{k=1}^{n} \sum_{l<k} \sum_{V_{a}=A, Z, W^{\pm}} \delta_{i_{k}^{\prime}
  i_{k} i_{l}^{\prime} i_{l}}^{V_{a}, \mathrm{SSC}}(k, l) \mathcal{M}_{0}^{i_{1}
  \ldots i_{k}^{\prime} \ldots i_{l}^{\prime} \ldots i_{n}}
,\end{align}
where the coefficient matrices for the different gauge bosons $V_a=A,Z,W^\pm$
are
\begin{align*}
  \delta_{i_{k}^{\prime} i_{k} i_{l}^{\prime} i_{l}}^{V_a, \mathrm{SSC}}(k, l)
  &=2 I_{i_{k}^{\prime} i_{k}}^{V_a}(k)
  I_{i_{l}^{\prime} i_{l}}^{\bar{V}_a}(l) \, \log \frac{\left|r_{k l}\right|}{s} l(s)
.\end{align*}
Note that while the photon couplings $I^A$ are diagonal, the eigenvalues for $I^Z$ and $I^\pm
\equiv I^{W^\pm}$ can be non-diagonal, leading again to replacements of external legs
as described in the LSC case.

\paragraph{\bf Collinear or soft single logarithms: C}
The two sources that provide either collinear or soft
logarithms are the running of the field renormalisation constants,
and the collinear limit of the loop diagrams where one external leg splits into
two internal lines, one of which being a vector boson $V_a$.
The ensuing correction factor can be written as a sum over external legs:
\begin{align}\label{eq:deltaC}
  \delta^{\mathrm{C}} \mathcal{M}^{i_{1} \ldots i_{n}}
  =\sum_{k=1}^{n} \delta_{i_{k}^{\prime} i_{k}}^{\mathrm{C}}(k)
  \mathcal{M}_{0}^{i_{1} \ldots i_{k}^{\prime} \ldots i_{n}}
.\end{align}
For chiral fermions $f$, the coefficient matrix is given by
\begin{equation}\label{eq:C_coeff_fermions}
  \delta_{f_{\sigma}f_{\sigma^{\prime}}}^{\mathrm{C}}\left(f^{\kappa}\right)
  =\delta_{\sigma
  \sigma^{\prime}}\left[\frac{3}{2}
  C_{f^{\kappa}}^{\mathrm{ew}}-\frac{1}{8
  s_{\mathrm{w}}^{2}}\left(\left(1+\delta_{\kappa \mathrm{R}}\right)
  \frac{m_{f_{\sigma}}^{2}}{M_W^{2}}+\delta_{\kappa \mathrm{L}}
  \frac{m_{f_{-\sigma}}^{2}}{M_W^{2}}\right)\right]
  l(s)
  = \delta_{f_{\sigma}f_{\sigma^{\prime}}}^{\mathrm{\overline{C}}}\left(f^{\kappa}\right)
  + \delta_{f_{\sigma}f_{\sigma^{\prime}}}^{\mathrm{Yuk}}\left(f^{\kappa}\right)
.\end{equation}
We label the chirality of the fermion by $\kappa$,
which can take either the value $L$ (left-handed) or $R$ (right-handed).
The label $\sigma$ specifies the isospin, its values $\pm$ refer to up-type
quarks/neutrinos or down-type quarks/leptons, respectively.
The sine (cosine) of the Weinberg angle is denoted by $s_w$ ($c_w$).
Note that we further subdivide the collinear contributions for fermion legs
into ``Yukawa'' terms formed by terms proportional to the ratio of the fermion masses
over the $W$ mass, which we denote by the superscript ``Yuk'', and the
remaining collinear contributions denoted by the superscript ``$\overline{\text{C}}$''.
These Yukawa terms only appear for external fermions, such that for all other
external particles $\varphi$, we have $\delta^\text{C}_{\varphi'\varphi} = \delta^{\overline{\text{C}}}_{\varphi'\varphi}$.
For charged and neutral transverse gauge bosons,
\begin{align}
  \delta_{W^{\sigma} W^{\sigma^{\prime}}}^{C}\left(V_{\mathrm{T}}\right)
  =
  \delta_{\sigma
  \sigma^{\prime}}\frac{1}{2} b_{W}^{\mathrm{ew}} l(s)
  \quad\text{and}\quad
  \delta_{N^{\prime}N}^{\mathrm{C}}\left(V_{\mathrm{T}}\right)
  =\frac{1}{2}\left[E_{N^{\prime} N}
  b_{A Z}^{\mathrm{ew}}+b_{N^{\prime} N}^{\mathrm{ew}}\right] l(s)
  \quad\text{with}\quad
  E\equiv\left(\begin{array}{cc}{0} & {1} \\ {-1} & {0}\end{array}\right)
\end{align}
must be used, where the $b^\text{ew}$ are combinations of Dynkin operators
that are proportional to the one-loop coefficients of the $\beta$-function
for the running of the gauge-boson self-energies and mixing energies.
Their values are given in terms of $s_w$ and $c_w$
in~\cite{Denner:2000jv}.

Longitudinally polarised vector bosons are again replaced with
Goldstone bosons.
When using the matrix element on the right-hand side
of Eq.~\eqref{eq:deltaC} in the physical EW phase,
the following (diagonal) coefficient matrices must be used
for charged and neutral longitudinal gauge bosons:
\begin{align}
  \delta_{W^{\sigma} W^{\sigma^{\prime}}}^{C}\left(V_{\mathrm{L}}\right)
  \to
  \delta_{\phi^{\pm} \phi^{\pm}}^{\mathrm{C}}(\Phi)
  &=\left[2 C_{\Phi}^{\mathrm{ew}}-\frac{N_{\mathrm{C}}}{4 s_{\mathrm{w}}^{2}} \frac{m_t^{2}}{M_W^{2}}\right] l(s)
  ,\\
  \delta_{N^{\prime}N}^{\mathrm{C}}\left(V_{\mathrm{T}}\right)
  \to
  \delta_{\chi \chi}^{\mathrm{C}}(\Phi)
  &=\left[2 C_{\Phi}^{\mathrm{ew}}-\frac{N_{\mathrm{C}}}{4 s_{\mathrm{w}}^{2}} \frac{m_t^{2}}{M_W^{2}}\right] l(s)
    ,
\end{align}
where $N_\text C=3$ is the number of colour charges.

\paragraph{\bf Parameter Renormalisation logarithms: PR}
The last contribution we consider is the one coming from the
renormalisation of EW parameters, such as boson and fermion
masses and the QED coupling $\alpha$, and all derived quantities. The way we
extract these terms, is by running all EW parameters up to a
given scale, $\mu_\text{EW}$, which in all cases presented here corresponds to the
partonic scattering centre of mass energy, and re-evaluate the matrix element
value with these evolved parameters. We then take the ratio of this `High
Energy' (HE) matrix element with respect to the original value, such
that, calling $\{p_{\text{ew}}\}$ the complete set of EW parameters,
\begin{equation}
  \label{eq:deltapr}
  \delta^\text{PR}_{i_1\dots i_n} = \left(\frac{\mathcal{M}_{\text{HE}}^{i_1\dots
      i_n}(\{p_{\text{ew}}\}(\mu_{\text{EW}}))}{\mathcal{M}^{i_1\dots
      i_n}(\{p_{\text{ew}}\})} - 1 \right) \sim
\frac{1}{\mathcal{M}^{i_1\dots i_n}}\sum_{p\in\{p_{\text{ew}}\}}\frac{\delta
  \mathcal{M}^{i_1\dots i_n}}{\delta p}\,\delta p.
\end{equation}
The evolution of each EW parameter is obtained
through
\begin{equation}
  p_{\text{ew}}(\mu_{\text{EW}}) =  \{p_{\text{ew}}\}\left( 1 + \frac{\delta
      p_{\text{ew}}}{p_{\text{ew}}} \right),
\end{equation}
and the exact expressions for $\delta p_{\text{ew}}$ can be found in
Eqs.~(5.4--5.22) of \cite{Denner:2000jv}.

The right hand side of Eq.~\eqref{eq:deltapr} corresponds to the
original derivation of Denner and Pozzorini, while the left hand side
is the actual implementation we have used. The two differ by terms
that are formally of higher order,
$(\alpha\,\log\mu_{\text{EW}}^2/m^2_W)^2$. In fact, although they are
logarithmically enhanced, they are suppressed by an additional power
of $\alpha$ with respect to the leading terms considered here,
$\alpha\,\log^2\mu_{\text{EW}}^2/m^2_W$.

\paragraph{\bf Generating event samples with EW Sudakov logarithmic contributions}
After having discussed the individual contributions $c$, we can
return to~Eq.~\eqref{eq:kfactor}, for which we now have all the
ingredients to evaluate it for an event with the phase-space point
$\Phi$. Defining the relative logarithmic contributions $\Delta^c$,
we can rewrite it as
\begin{equation}
  \label{eq:kfactor_contribs}
  K_{\text{NLL}}\left(\Phi\right)
  = 1 + \sum_c \Delta^c = 1 + \Delta^{\overline{\text{LSC}}} + \Delta^\text{Z} + \Delta^\text{SSC} + \Delta^{\overline{\text{C}}} + \Delta^\text{Yuk} + \Delta^\text{PR}.
\end{equation}
In the event sample, the relative contributions $\Delta^c$ are given in the form of named
\HepMC weights~\cite{Dobbs:2001ck} (details on the naming will be given in the
manual of the upcoming \Sherpa release).
This is done to leave the user freedom on how to combine such weights with the
main event weight.

In the context of results, Sec.~\ref{sec:examples-testin}, we employ these corrections in either of two
ways. One way is to include them at fixed order,
\begin{equation}
  \label{eq:kfactor_fixed_order}
  \text{d}\sigma^\text{LO + NLL}\left(\Phi\right) =
  \text{d}\Phi\,\mathcal{B}\left(\Phi\right) \, K_{\text{NLL}}\left(\Phi\right)\, ,
\end{equation}
where $\mathcal{B}$ is the Born contribution.
The alternative is to exploit the fact that Sudakov EW logarithms
exponentiate (see Refs.~\cite{Fadin:1999bq}--\cite{Fuhrer:2010eu}) to
construct a resummed fully differential cross section
\begin{equation}
  \label{eq:kfactor_resum}
  \text{d}\sigma^\text{LO + NLL (resum)}\left(\Phi\right) =
  \text{d}\Phi\, \mathcal{B}\left(\Phi\right) \,   K^\text{resum}_{\text{NLL}}\left(\Phi\right)
  = \text{d}\Phi\, \mathcal{B}\left(\Phi\right) \,e^{\left(1 - K_\text{NLL}\left(
        \Phi \right)\right)}\, ,
\end{equation}
following the approach discussed in~\cite{Lindert:2017olm}.

\paragraph{\bf Matching to higher orders and parton showers} \Sherpa
internally provides the possibility to the user to 
obtain NLO corrections, both of
QCD~\cite{Gleisberg:2007md} and EW~\cite{Schonherr:2017qcj} origin,
and to further generate fully showered~\cite{Schumann:2007mg,Hoche:2015sya}
and hadronised events~\cite{Bothmann:2019yzt}.
In addition one is able to merge samples with higher multiplicities in QCD
through the CKKW~\cite{Catani:2001cc}, or the
\MEPSatNLO~\cite{Hoeche:2012yf} algorithms.
In all of the above cases (except the NLO EW), the corrections implemented here can be
simply applied
using the K factor methods in \Sherpa
to one or all the desired processes of the
calculations, as there is no double counting between EW Sudakov
corrections and pure QCD ones. A similar reasoning can be applied
for the combination with a pure QCD parton shower.
Although we do not report the result of this
additional check here, we have tested the combination of the EW 
Sudakov corrections with the default shower of \Sherpa.
Technical checks and physics applications for the combination with
matching and merging schemes, and for the combination with QED
logarithms, are left for a future publication.

If one aims at matching Sudakov logarithms to higher-order EW effects, such as
for example combining fixed order NLO EW results with a resummed NLL Sudakov
correction, it is for now required for the user to manually do the subtraction
of double counting that one encounters in these cases.
An automation of such a scheme is also outside the scope of this publication,
and will be explored in the future.

\section{Results}
\label{sec:examples-testin}

Before discussing our physics applications, we report a number of exact
comparisons to other existing calculations of NLL EW Sudakov logarithms.
A subset of the results used for this comparison for $pp \to Vj$ processes is
shown in App.~\ref{app:vj}.
They all agree with reference ones on a sub-percent level over the entire
probed transverse momentum range from $p_{T,V}=\SI{100}{\GeV}$ to
\SI{2000}{\GeV}.
In addition to this, the implementation has been guided by passing a number of
tests based on a direct comparison of tabulated numerical values for each
contribution discussed in Sec.~\ref{sec:imple} in the high-energy limit,
that are given for several electron-positron collision processes
in~\cite{Denner:2000jv} and for the $pp \to WZ$ and $pp \to W\gamma$ processes
in~\cite{Pozzorini:2001rs}.
Our final implementation passes all these tests.

In the remainder of this section, we present a selection of physics results
obtained using our implementation and where possible show comparisons with
existing alternative calculations.
The aim is twofold, first we wish to show the
variety of processes that can be computed with our implementation, and
second we want to compare to existing alternative methods to obtain EW
corrections to further study the quality of the approximation.
In particular, where available, we compare our NLL predictions to full NLO EW
corrections, as well as to the \EWvirt approximation defined
in~\cite{Kallweit:2015dum}.
It is important to note that we only consider EW corrections to purely
EW processes here, such that there are no subleading Born corrections
that might otherwise complicate the comparison to a full EW NLO calculation or
to the \EWvirt approximation.
We consider diboson production, dijet production, and $Z$ production in
association with 4 jets.
In all cases, we focus on the discussion of (large) transverse momentum
distributions, since they are directly sensitive to the growing effect of the
logarithmic contributions when approaching the high-energy limit.

The logarithmic contributions are applied to parton-level LO calculations provided by the \Comix
matrix element generator implemented in \Sherpa.
The parton shower and non-perturbative parts of the simulation are disabled,
including the simulation of multiple interactions and beam remnants,
and the hadronisation.
Also higher-order QED corrections to the matrix element and by YFS-type resummation are turned off.
The contributions to the NLL corrections as defined in Sec.~\ref{sec:imple} are
calculated individually and combined a-posteriori, such that we can study their
effects both individually and combined.
We consider the fixed-order and the
resummed option for the combination, as detailed in Eqs.~\eqref{eq:kfactor_fixed_order}
and~\eqref{eq:kfactor_resum}, respectively.

For the analysis, we use the Rivet 2~\cite{Buckley:2010ar} framework,
and events are passed to analysis using
the \HepMC event record library.
Unless otherwise specified, simulations are obtained using the
NNPDF3.1, next-to-next-to-leading order PDF set~\cite{Ball:2017nwa},
while in processes where we include photon initiated processes we instead use
the NNPDF3.1 LUX PDF set~\cite{Bertone:2017bme}. In all
cases, PDFs are obtained through the LHAPDF~\cite{Buckley:2014ana}
interface implemented in \Sherpa.
When jets appear in the final state, we cluster them
using the anti-$k_T$
algorithm~\cite{Cacciari:2008gp} with a jet radius parameter of
$R=0.4$, through an interface to \Fastjet\cite{Cacciari:2011ma}.
The CKM matrix in our calculation is equal to the unit matrix,
i.e.\ no mixing of the quark generations is allowed.
Electroweak parameters are determined using tree-level relations using a QED
coupling value of $\alpha(m_Z) = 1 / 128.802$ and the following set of masses
and decay widths, if not explicitly mentioned otherwise:
\begin{align*}
  m_W&=\SI{80.385}{\GeV} &
  m_Z&=\SI{91.1876}{\GeV} &
  m_h&=\SI{125}{\GeV} \\
  \Gamma_W&=\SI{2.085}{\GeV} &
  \Gamma_Z&=\SI{2.4952}{\GeV} &
  \Gamma_h&=\SI{0.00407}{\GeV}.
\end{align*}
Note that $\alpha$ is not running in the nominal calculation
as running effects are all accounted for by the PR contributions.

\subsection{$WW$ production in $pp$ collisions at 13 and 100 TeV}

Our first application is the calculation of EW Sudakov effects in on-shell $W$
boson pair production at hadron colliders, which has lately been experimentally
probed at the \LHC~\cite{Sirunyan:2017bey,Aaboud:2019nkz}, e.g.\ to search for
anomalous gauge couplings.
We compare the Sudakov EW approximation to both the full NLO EW calculation and
the \EWvirt approximation.
The latter has been applied to $WW$ production in
\cite{Kallweit:2017khh,Brauer:2020kfv}.
In addition to that, EW corrections for $WW$ production have also been studied
in~\cite{Kuhn:2011mh,Bierweiler:2012kw,Baglio:2013toa,Gieseke:2014gka,Li:2015ura,Biedermann:2016guo},
and NNLL EW Sudakov corrections have been calculated in~\cite{Kuhn:2011mh}.

We have performed this study at
current \LHC energies, $\sqrt s = \SI{13}{\TeV}$, as 
well as at a possible future hadron collider with $\sqrt s = \SI{100}{\TeV}$.
In all cases, we include photon induced channels
(they can be sizeable at large energies \cite{Bierweiler:2012kw}), and we set the
renormalisation and factorisation scales to $\mu_{R,F}^2 =
\frac{1}{2}( 2m_W^2 + \sum_i p^2_{T,i})$,
where the sum runs over the final-state $W$ bosons,
following the choice for gauge-boson pair production in~\cite{Accomando:2001fn}.
Lastly, we set the widths of the $Z$ and $W$ boson consistently to 0,
as the $W$ is kept on-shell in the matrix elements.
\begin{figure}[!t]
    \centering
    \includegraphics[width=0.49\linewidth]{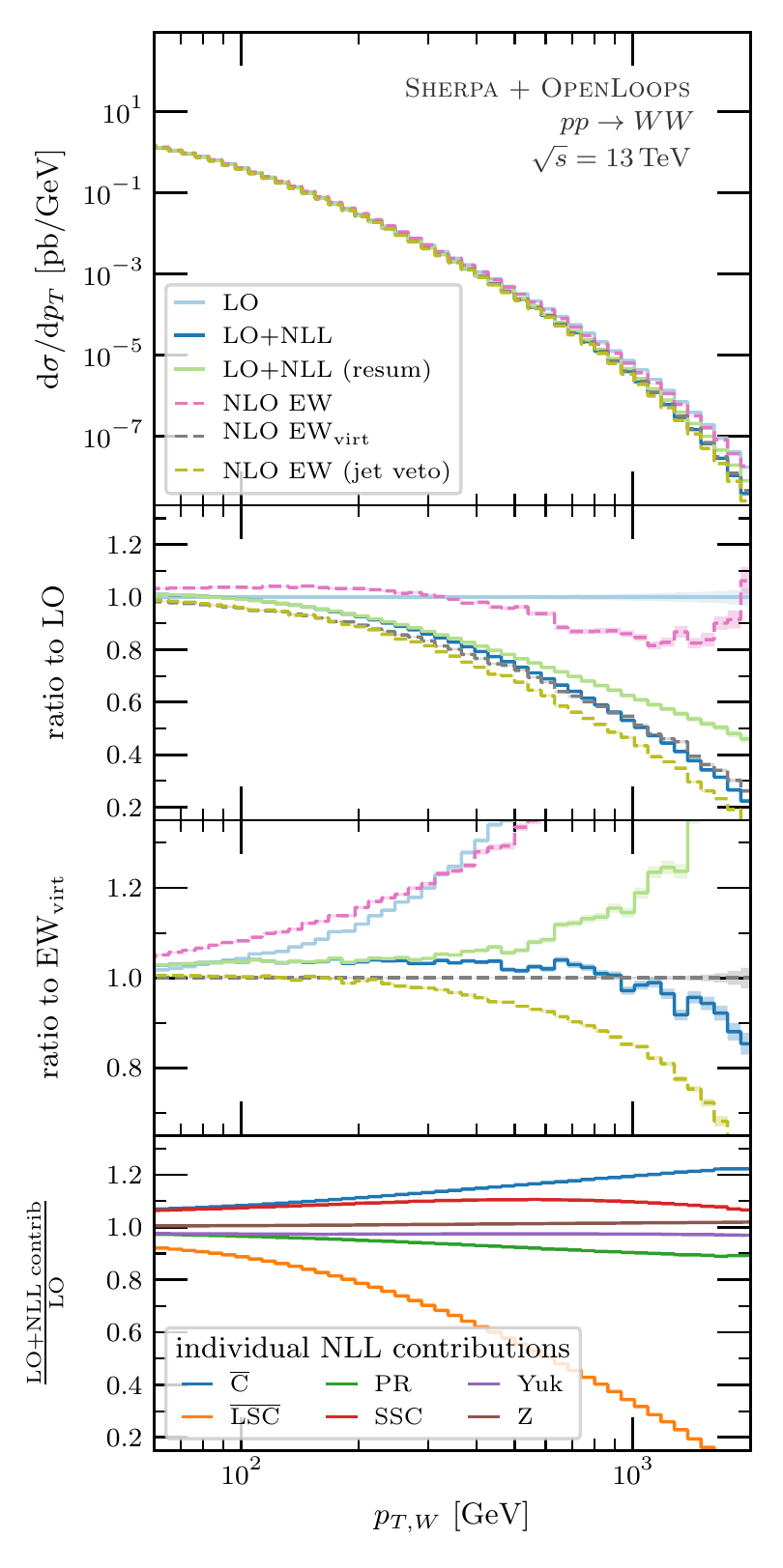}
    \hfill
    \includegraphics[width=0.49\linewidth]{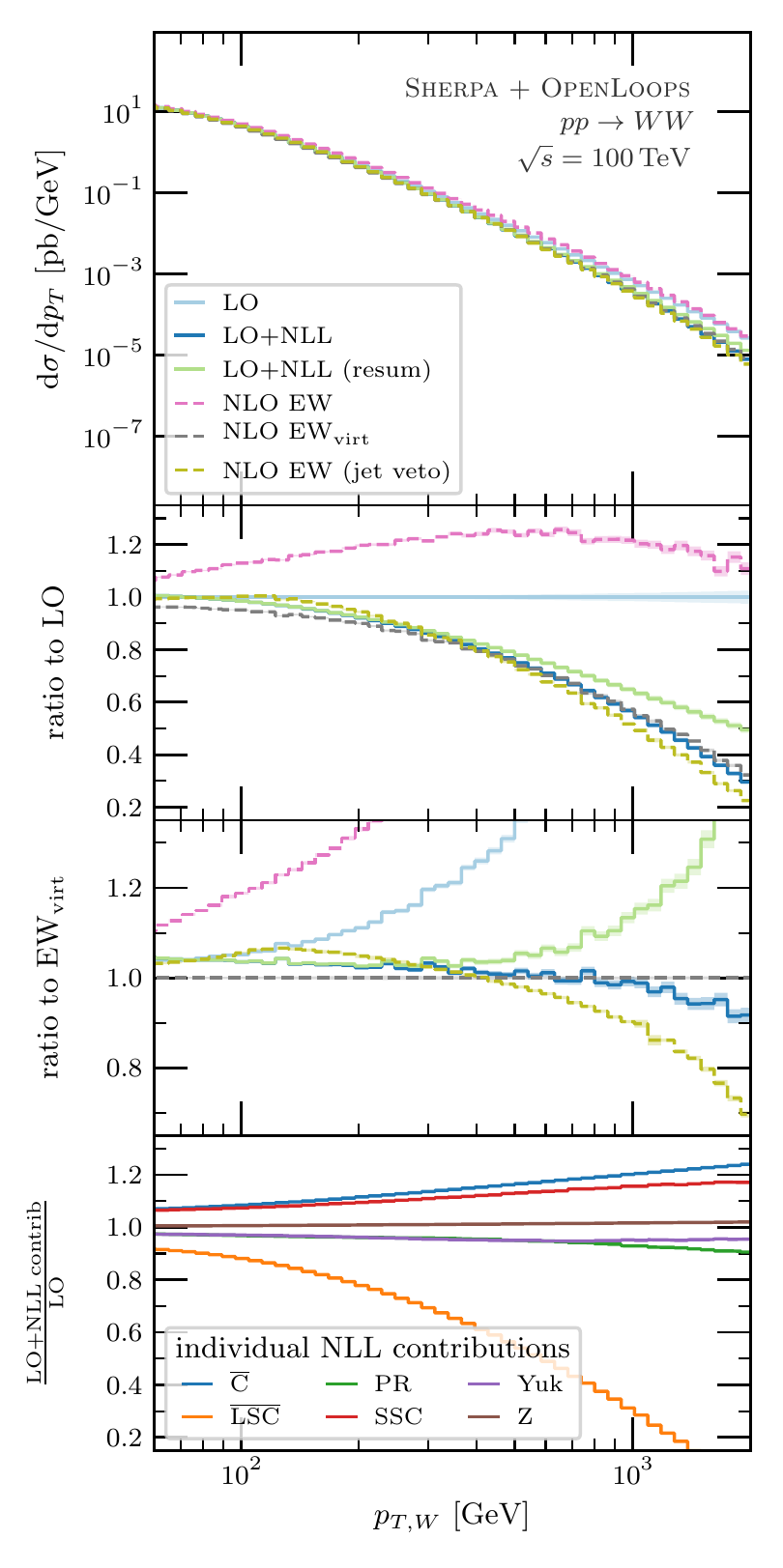}
    \caption{The transverse momentum of the
    individual $W$ bosons in $W$ boson pair production
    from proton-proton collisions (including photon induced channels)
    at $\sqrt s = \SI{13}{\TeV}$ (left) and $\SI{100}{\TeV}$ (right).
    The baseline LO and NLO EW calculations are compared with the results of
    the LO+NLL calculation and its variant, where the logarithmic corrections
    are resummed (``LO+NLL (resum)'').
    In addition, the virtual approximation \EWvirt and a variant of
    the NLO EW calculation with additional jets vetoed are also included.
    The ratio plots show the ratios to the LO and to the \EWvirt calculation,
    and the relative size of each NLL contribution.}
    \label{fig:ww}
\end{figure}

In Fig.~\ref{fig:ww} we report results for the transverse momentum of
each $W$ boson in the final state for both centre of
mass energies. Plots are divided into four panels, of which the first
one collects results for the various predictions. The second and third
panels report the ratio to the leading order and to the \EWvirt
approximation, respectively. The aim is to show the general behaviour
of EW corrections in the tail of distributions in the former, while
the latter serves as a direct comparison between the Sudakov and
the \EWvirt approximations. Finally the fourth panel shows the
relative impact of the individual contributions $\Delta^c$ appearing in
Eq.~\eqref{eq:kfactor_contribs}.

Looking at the second panel, we see that all but the full NLO EW
calculation show a strong suppression in the $p_T > m_W$ region,
reaching between $\SI{-70}{\percent}$ and $\SI{-90}{\percent}$ at
$p_T=\SI{2}{\TeV}$. This effect is the main effect we discuss in this work,
and it is referred to as Sudakov suppression. To explicitly confirm that
this behaviour originates from virtual contributions of EW nature, we
compare the Sudakov LO+NLL curve to the NLO EW and \EWvirt
approximations. Indeed, the latter only takes into account virtual
corrections and the minimal amount of integrated counter terms to
render the cross section finite.
The Sudakov approximation is close to that for both centre of mass energies
(see also the third panel), with deviations of the order of a few percent.
It begins to deviate more at the end of the spectrum, with similar
behaviours observed for both collider setups.
However, with a Sudakov suppression of $\SI{70}{\percent}$ and more, we are
already in a regime where the relative corrections become $\mathcal{O}(1)$
and a fixed logarithmic order description becomes invalid.
In fact, at $p_{T,W} \gtrsim \SI{3}{\TeV}$, the LO+NLL becomes negative
both at $\sqrt s = \SI{13}{\TeV}$ and \SI{100}{\TeV},
the same being true for the \EWvirt approximation.
Note in particular, that this is the main reason for choosing to show
the $p_T$ distribution only up to $p_T = \SI{2}{\TeV}$ in
Fig.~\ref{fig:ww} for both collider setups.

It is also clear from the second panel that for both setups, the full NLO EW
calculation does not show such large suppressions, and in the context
of the question whether to use EW Sudakovs or the \EWvirt to approximate the full
corrections, this may be worrisome. However, in the full calculation
we have included the real emission matrix elements, which also show a
logarithmic enhancement at high $p_T$, as e.g.\ discussed in~\cite{Baglio:2013toa}.
In this case, the real emission contribution almost
entirely cancels the Sudakov suppression at $\sqrt{s}=\SI{13}{\TeV}$,
and even overcompensates it by about \SI{20}{\percent} in the
$p_{T,W}\lesssim\SI{1}{\TeV}$ region for $\sqrt{s}=\SI{100}{\TeV}$.
To show that this is the case, we have also reported a jet-vetoed
NLO EW simulation, which indeed again shows the high $p_T$
suppression as expected.
Moreover, we have included a prediction labelled ``LO+NLL (resum)'', where we
exponentiate the Sudakov contribution using Eq.~\eqref{eq:kfactor_resum}.
It is similar to the NLL approximation, but
resumming these logarithms leads to a smaller suppression in the
large $p_T$ tail, which is reduced by about \SI{20}{\percent}
at $p_T=\SI{2}{\TeV}$, thus increasing the range of validity
compared to the non-exponentiated prediction.
Moreover, this agrees well qualitatively with the NNLL result reported
in~\cite{Kuhn:2011mh}, and suggests that even higher-order logarithmic effects
should be rather small in comparison in the
considered observable range.

Lastly, we compare the individual LL and NLL contributions. As expected, we
find that the double logarithmic $\overline{\text{LSC}}$ term is the
largest contribution and drives the Sudakov suppression. 
Some single logarithmic terms, in particular the
$\overline{\text{C}}$ terms, also give a sizeable
contribution, reducing the net suppression. 
This confirms that the inclusion of single logarithmic terms is needed in
order to provide accurate predictions in the Sudakov approximation,
with deviations of the order of $\mathcal{O} (\SI{10}{\percent})$ with
respect to the \EWvirt calculation.
Comparing the individual contributions for the two collider energies,
we see qualitatively similar effects.
As can be expected from a larger admixture of $b\bar b$ initial states,
the $\Delta^\text{Yuk}$ is enhanced at the larger collider energy.

\subsection{EW-induced dijet production in $pp$ collisions at 13 TeV}

For the second comparison in this section, we simulate purely
EW dijet production in hadronic collisions at $\sqrt s = \SI{13}{\TeV}$ at
the Born-level perturbative order $\mathcal{O}(\alpha^2\alpha_s^0)$.
As for the case of diboson production we add photon initiated
channels, and we also include photons in the set of final-state partons,
such that $\gamma\gamma$ and $\gamma j$ production is also part of our sample.
Partons are thus clustered into jets which are then sorted by their $p_T$.
We select events requiring at least two jets, the leading jet (in
$p_T$) to have a $p_T >
\SI{60}{\GeV}$ and the subleading jet to have a $p_T > \SI{30}{\GeV}$. Note in
particular that for all but the real emission case in the full NLO EW
simulation, this corresponds to imposing a $p_T$ cut on the two
generated partons.
The renormalisation and factorisation scales are set to $\mu_{R,F}^2 = \hat{H}^2_T = (\sum_i
p_{T,i})^2$, where the sum runs over final-state partons.

We compare our LO+NLL EW results, as in the previous subsection, with the LO, the NLO EW and
the EW$_\text{virt}$ predictions. NLO EW corrections for dijet
production have been first discussed
in~\cite{Dittmaier:2012kx,Frederix:2016ost},
while~\cite{Reyer:2019obz} discusses those corrections in the context
of the 3-to-2-jet ratio $R_{32}$.
In this context, we only consider EW corrections to the Born process
described above, i.e.\ we consider the
$\mathcal{O}(\alpha_s^0\alpha^3)$ contributions,
which is a subset of the contributions considered in the above references.

In Fig.~\ref{fig:ew_dijets_and_Z4jets} (left), we present the transverse
momentum distribution of the leading jet $p_T$ given by the LO+NLL EW
calculation, again both at fixed order and resummed, see
Eqs.~\eqref{eq:kfactor_fixed_order} and~\eqref{eq:kfactor_resum}.
As for $W$ pair production, the plot is divided into four panels.
However, this time we do not include jet-vetoed NLO results, as in this case we
do not observe large real emission contributions.
The panels below the main plot give the ratio to the LO calculation, the ratio
to the \EWvirt approximation, and the ratio of each NLL contribution to the LO
calculation.

Compared to $W$ pair production, we observe a smaller but still sizeable
Sudakov suppression,
reaching approximately $-\SI{40}{\percent}$ at $p_T=\SI{2}{\TeV}$.  
The NLO contributions not included in the NLO EW$_\text{virt}$ (i.e.\  mainly
real emission terms) are small and cancel the Sudakov suppression only by a few
percent and as such both the Sudakov and the \EWvirt approximations agree well
throughout the $p_T$ spectrum with the full NLO calculation.
The same is true for the resummed NLL case which gives a Sudakov suppression of
$\SI{-30}{\percent}$ for $p_T=\SI{2}{\TeV}$.
Note that a small step can be seen for $p_T\sim m_W$ in the Sudakov
approximation. The reason for this is that we force, during the
simulation, all Sudakov contributions to be zero when at least one of the
invariants formed by the scalar product of the external momenta is
below the $W$ mass, as Sudakov corrections are technically only valid in the high-energy
limit.
This threshold behaviour can be disabled, giving a smoother
transition between the LO and the LO+NLL, as can be seen for example
in Fig.~\ref{fig:ww}.

Comparing the individual NLL contributions we find again that the
double logarithmic $\overline{\text{LSC}}$ term is the largest
contribution, its size is however reduced by a third compared
to the diboson case, since the prefactor $C^\text{ew}$ in
Eq.~\eqref{eq:LSC_coeff} is smaller for quarks and photons compared to $W$
bosons.
Among the single logarithmic terms the $\overline{\text{C}}$ and PR terms are
the most sizeable over the whole range, and are of a similar size compared to the
diboson case, while the SSC contribution only give a small contribution.
Again, subleading terms must be included to approximate the NLO calculation at
the observed \SI{10}{\percent} level, although we observe an almost
entire accidental cancellation, in this case, of the SL terms.

\subsection{Off-shell $Z$ production in association with 4 jets in $pp$ collisions at 13 TeV}
\label{neat-example}

As a final example, we present for the first time
the LO+NLL calculation for $e^+ e^-$ production in association with
four additional jets 
from proton-proton collisions at $\sqrt s = \SI{13}{\TeV}$.
The process is one of the key benchmark processes at the \LHC to
make precision tests of perturbative QCD
and is a prominent background constituent for several Standard Model
and New Physics processes.
In this case, we neglect photon induced contributions, to better compare to the
NLL effects in $Z$ plus one-jet production presented in App.~\ref{app:vj},
which in turn is set up as a direct comparison to~\cite{Kuhn:2004em}. 
For the same reason, in this case, we only consider QCD partons in the
final state.
The factorisation and renormalisation scale are set to $\mu_F^2 =
\mu_R^2 = \hat{H}_T^{\prime 2} = (M_{T,\ell\ell} + \sum_i p_{T,i})^2/4.0$, where
$M_{T,\ell\ell}$ is the transverse mass of the electron-positron
pair and the sum runs over the final-state partons. This choice is
inspired by~\cite{Anger:2017nkq}, where the full next-to-leading QCD
calculation is presented.

\begin{figure}[!b]
    \centering
    \includegraphics[width=0.49\linewidth]{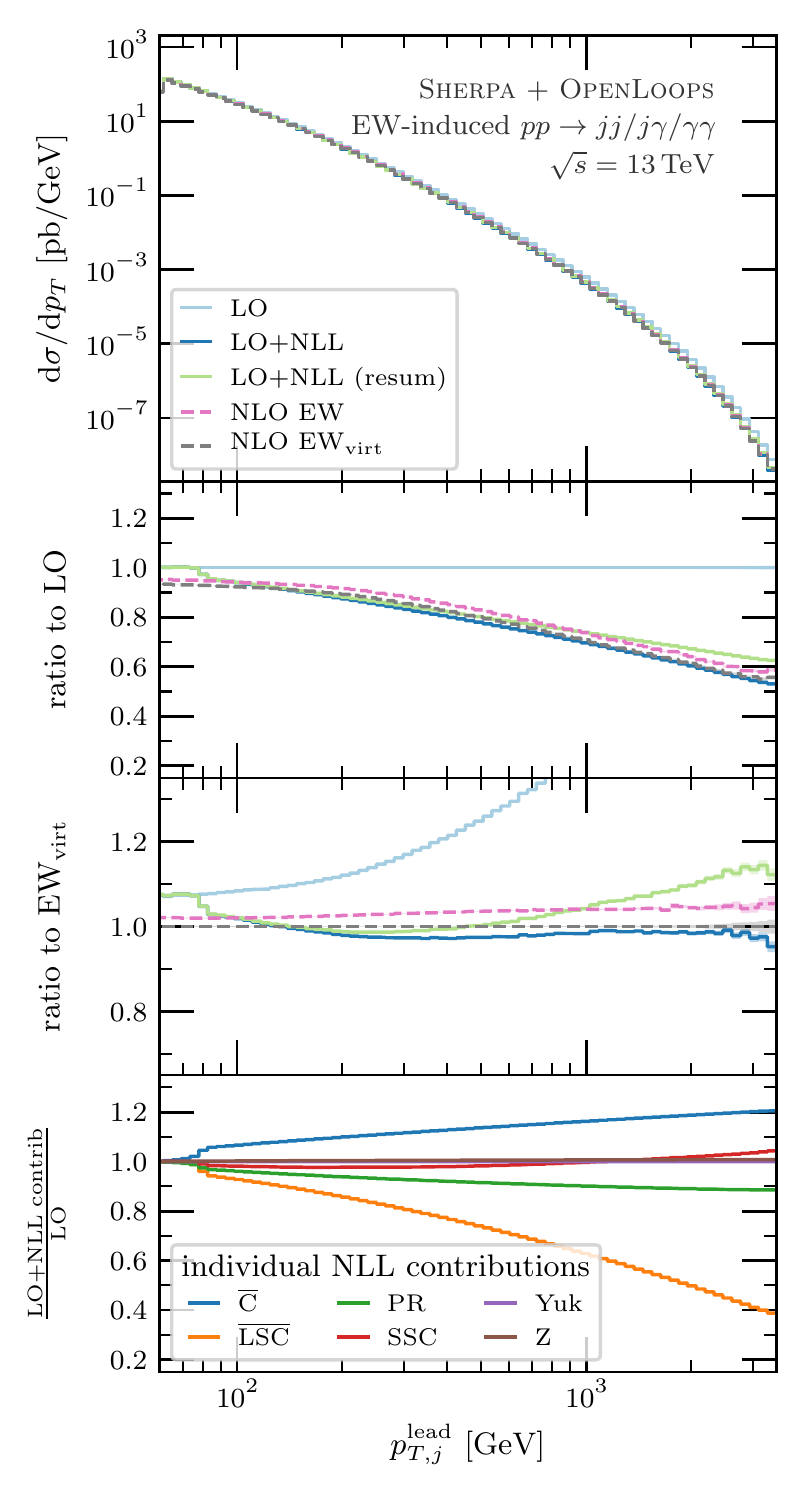}
    \hfill
    \includegraphics[width=0.50\linewidth]{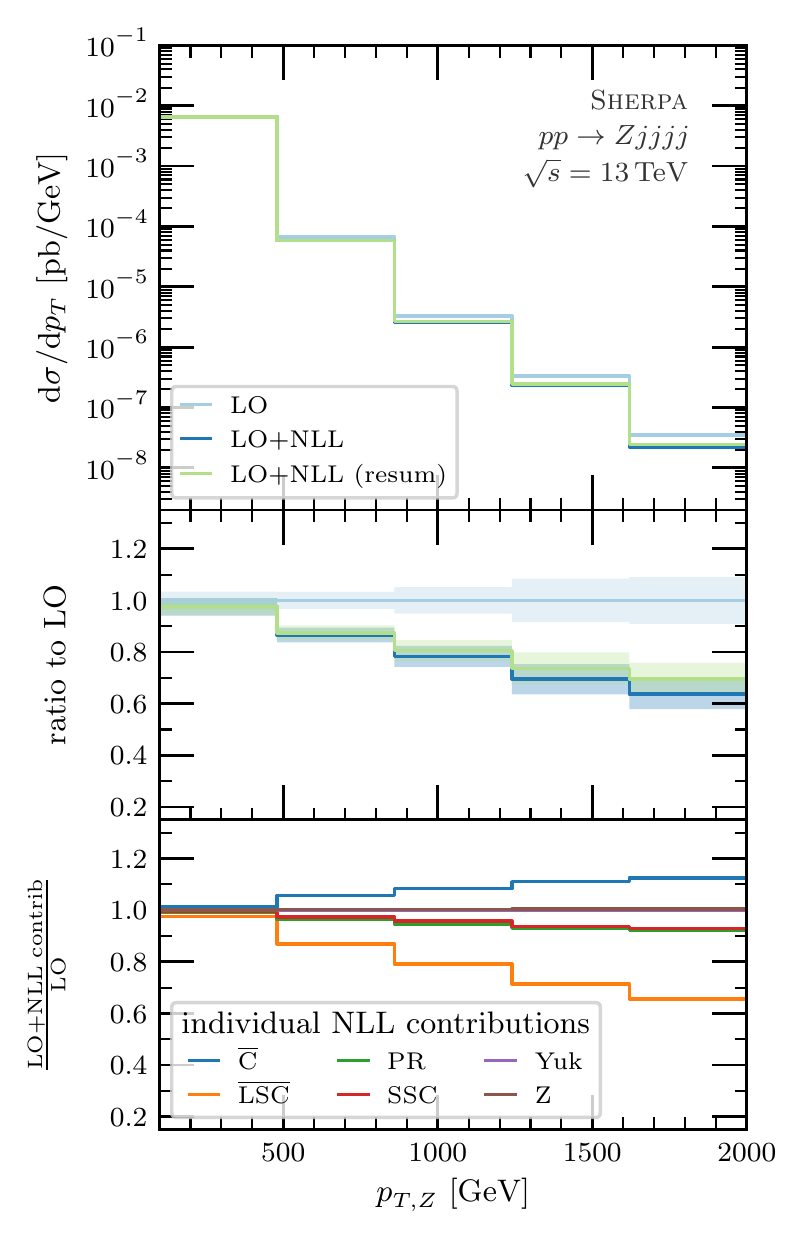}
    \caption{The transverse momentum of the
      leading jet in EW-induced dijet production
    in proton-proton collisions (including photon channels),
    and for the reconstructed $Z$ boson in $e^+ e^-$ plus four jets production,
    For the dijet production, LO and NLO calculations are shown,
    whereas for the $Z$ plus jets production only the LO is shown.
    These baseline calculations are compared with the results of the LO+NLL calculation,
    both at fixed-order and resummed.
    In the dijet case, the virtual approximation \EWvirt is shown in addition.
    The ratio plots show the ratios to the LO and the \EWvirt calculations,
    and the relative size of each NLL contribution.}
    \label{fig:ew_dijets_and_Z4jets}
\end{figure}

As we consider here an off-shell $Z$, the invariant mass formed by its
decay products is on average only slightly above the $W$ mass threshold.
This may cause an issue as this is one of the invariants considered
in the definition of the high-energy limit discussed in Sec.~\ref{sec:imple},
and therefore this limit is strictly speaking not fulfilled.
In turn, this can introduce sizeable logarithms in particular in
Eq.~\eqref{eq:deltaSSC}. 
However, in practice we only see a small number of large $K$ factors,
that only contribute a negligible fraction of the overall cross
section. 
For set-ups at a larger centre of mass energy, one should monitor this
behaviour closely, as the average value
of $s$ in Eq.~\eqref{eq:deltaSSC} would increase too. 
We therefore foresee that a more careful treatment might be required
then, e.g.\ by vetoing EW Sudakov K factors whenever $|r_{kl}| \ll
s$.

In Fig.~\ref{fig:ew_dijets_and_Z4jets} (right), we present the LO+NLL
EW calculation of the transverse momentum distribution of the
reconstructed $Z$ boson.
To our knowledge, there is no existing NLO EW calculation to compare it
against\footnote{Note that while such a calculation has not yet been
  published, existing tools, including \Sherpa in
  conjunction with OpenLoops~\cite{Buccioni:2019sur}, are in
  principle able to produce such a set-up.},
hence we do not have a
ratio-to-NLO plot in this case. However, we do show the ratio-to-NLL
plot and the plot that shows the different contributions of the LO+NLL
calculation, as for the other processes. 

The sizeable error bands give the MC errors of the LO calculation.
Note that the errors of the LO and the LO+NLL calculation are fully
correlated,
since the NLL terms are completely determined by the phase-space point
of the LO event, and the same LO event samples are used for both
predictions.
Hence the reported ratios are in fact very precise.
With the aim of additionally making a comparison to the
$Z+j$ result reported in Fig.~\ref{fig:vj} (right panel), we opt for a linear
$x$ axis, in contrast with the other results of this section.
In both cases we see a similar overall LO+NLL effect, reaching
approximately $\SI{-40}{\percent}$
for $p_T\lesssim\SI{2}{\TeV}$. This in turn implies that the effects of
considering an off-shell $Z$ (as opposed to an on-shell decay) as well
as the additional number of jets have very little effect on the
size of the Sudakov corrections. 
Finding similar EW corrections for
processes that only differ in their number of additional QCD emissions
can be explained by the fact that the sum of EW charges of the
external lines are equal in this case. 
As has recently been noted in~\cite{Brauer:2020kfv}, this can be
deduced from the general expressions for one-loop corrections
in~\cite{Denner:2000jv} and from soft-collinear effective
theory~\cite{Chiu:2008vv,Chiu:2009ft}. 
Although the overall effect for $Zj$ and $Z+4j$ is found to be
very similar here, the individual contributions partly exhibit a
different behaviour between the two, with the SSC terms becoming
negative in the four-jet case and thus switching sign, and
the C terms becoming a few percent smaller.
It is in general noticeable that the SSC terms exhibit the strongest shape
differences among all processes considered in this study.
Finally, similarly to the previous studied cases, the resummed result gives a slightly reduced
Sudakov suppression, reaching approximately $\SI{-30}{\percent}$ for
$p_T\lesssim\SI{2}{\TeV}$, implying that in this case, higher
logarithmic contributions should be small.

\section{Conclusions}
\label{sec:conclusions}

We have presented for the first time a complete,
automatic and fully general implementation of the algorithm presented
in~\cite{Denner:2000jv} to compute double and single logarithmic
contributions of one-loop EW corrections, dubbed as EW Sudakov
logarithms.
These corrections can give rise to large shape distortions in the
high-energy tail of distributions, and are therefore an important
contribution in order to improve the accuracy of
the prediction in this region. Sudakov logarithms can provide a good
approximation of the full next-to-leading EW corrections.
An exponentiation of the corrections
can be used to resum the logarithmic effects and extend the
region of validity of the approximation.

In our implementation, each term contributing to the
Sudakov approximation is returned to the user in the form of an additional
weight, such that the user can study them individually, add their sum
to the nominal event weight, or exponentiate them first.
Our implementation will be made available with the
next upcoming major release of the \Sherpa Monte Carlo event
generator. We have tested our implementation against an array of
existing results in the same approximation, and for a variety of
processes against full NLO EW corrections and
the virtual approximation \EWvirt, which is also available
in \Sherpa. A selection of such tests
is reported in this work, where we see that indeed EW Sudakov
logarithms give rise to large contributions and model the full
NLO corrections well when real emissions are small.
We stress that our implementation is
not limited to the examples shown here, but it automatically computes
such corrections for any Standard Model process.
In terms of final-state multiplicity it
is only limited by the available computing resources.

In a future publication we plan to apply the new implementation in the context
of state-of-the-art event generation methods, in particular to combine them
with the matching and merging of higher-order QCD corrections and the QCD
parton shower, while also taking into account logarithmic QED corrections using
the YFS or QED parton shower implementation in \Sherpa.
This will allow for an automated use of the method
for the generation of event samples in any \LHC or future
collider context, in the form of optional additive weights.
As discussed towards the end of Sec.~\ref{sec:imple}, we foresee that this is a
straightforward exercise, since there is no double counting among the different
corrections, and applying differential K factors correctly within these methods
is already established within the \Sherpa framework.
The result should also allow for the inclusion of subleading Born contributions,
as the \EWvirt scheme in \Sherpa does.
The automated combination of (exponentiated) EW Sudakov logarithms with
fixed-order NLO EW corrections is another possible follow-up, given the
presence of phenomenologically relevant applications.

\section*{Acknowledgements}
We wish to thank Stefano Pozzorini for the help provided at various
stages with respect to his original work, as well as Marek Sch\"onherr,
Steffen Schumann, Stefan H\"oche
and Frank Krauss and all our \Sherpa colleagues for stimulating
discussions, technical help and comments on the draft.
We also thank Jennifer Thompson for the
collaboration in the early stage of this work. The work of DN
is supported by the ERC Starting Grant 714788 REINVENT.

\clearpage
\appendix
\numberwithin{equation}{section}
\setcounter{equation}{0}


\section{Validation plots for $V$ plus jet production in $pp$ collisions at 13 TeV}

\begin{figure}[h!]
    \centering
    \includegraphics[width=1.00\linewidth]{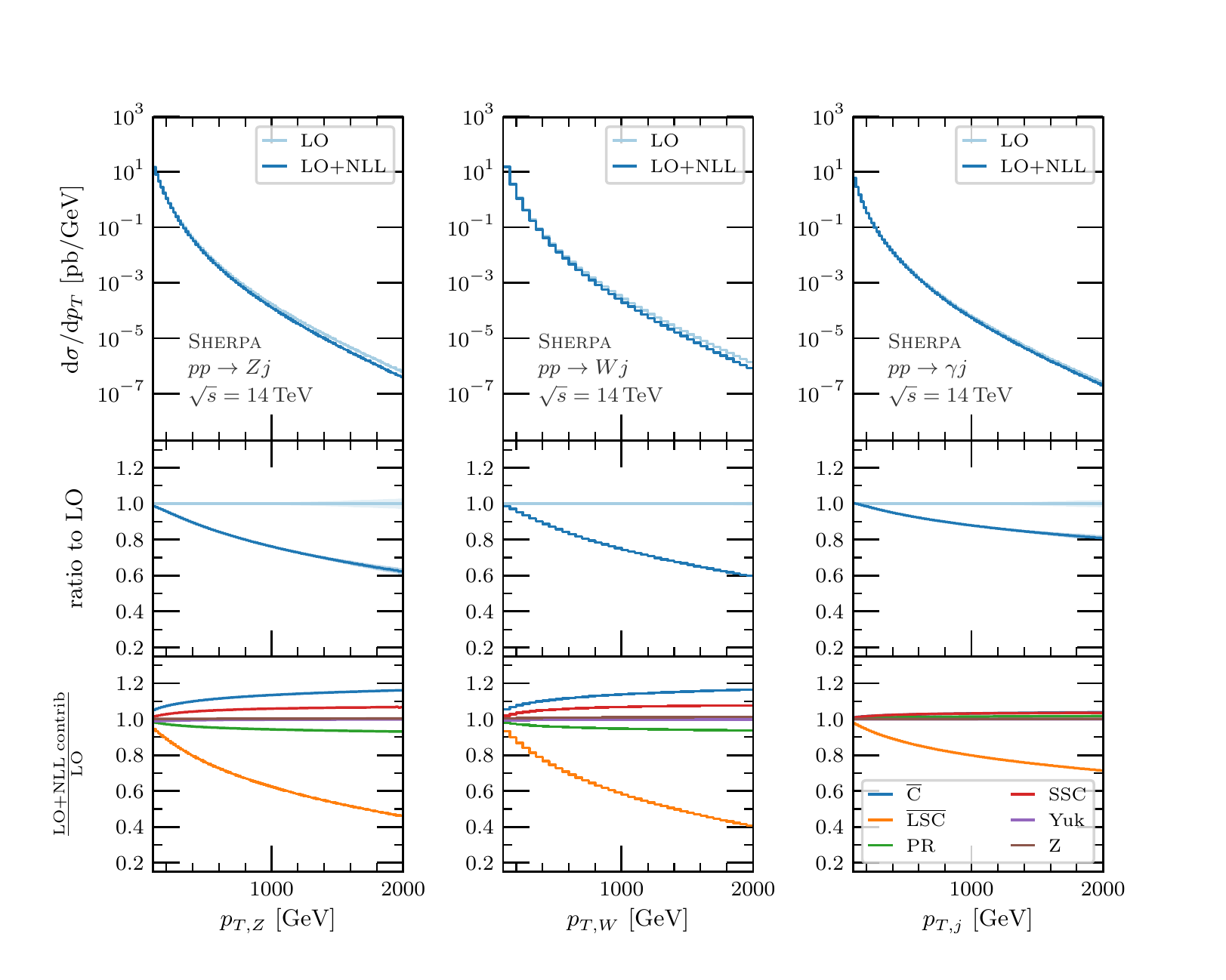}
    \caption{The transverse momentum of the vector boson in $Vj$ production for $V=W^+,Z,\gamma$.
    The baseline LO calculations are compared with the results of the LO+NLL calculation.
    The ratio plots show the ratios to the LO calculation,
    and the relative size of each NLL contribution.}
    \label{fig:vj}
\end{figure}

As an additional validation for our implementation,
we show in Fig.~\ref{fig:vj} the LO+NLL calculation
for $Z$, $W^+$ and $\gamma$ production, each in association with one jet.
The setups, the analysis, and the binning are configured to closely match the choices made in~\cite{Kuhn:2004em,Kuhn:2007cv,Kuhn:2005gv}, respectively, and the reader is referred to these publications for further details.
In particular, the photon is not considered to be part of the parton content of the incoming protons, and the MRST01 LO PDF set has been used~\cite{Martin:2002dr}.
We have confirmed that our NLL predictions agree with the reference NLL ones at a sub-percent level over the whole range of the observables.

\label{app:vj}

\label{app-1}
\clearpage
\bibliographystyle{amsunsrt_modp}
\bibliography{bib/journal}
\end{document}